# Approximation of the Masses and other properties of Unknown Hyperons in Standard Model

Imran Khan
*Department of Physics, University of Science and Technology, Bannu 28100, Pakistan*

Email address: immarwat@yahoo.com

***Abstract***—In particle physics, study of the symmetry and its breaking play very important role in order to get useful information about the nature. The classification and arrangements of subatomic particles is also necessary to study particle physics. Particles which are building blocks of nature are quarks, gluons and leptons. Baryons and Mesons composed of quarks were arranged by Gell-Mann and Okubo in their well-known Eight-Fold way up to SU(3) symmetry. Standard model of particles is composed of these particles. Particles in SU(4) also make some beautiful patron. These make some multiplets. but all the baryons with spin $J^P= 3/2^+$ in these multiplets have not been observed till date. In this paper the SU(4) multiplets have been organized and studied in an easy and new way. As a result we obtained some clues about the masses and other characteristics of the unknown hyperons. These approximations about the characteristics of the unidentified baryons have been recorded in this article. Mass formula for the baryon SU(4) multiplets have been obtained.

## I. Introduction

Baryons are made of three quarks ($qqq$). The three flavors up $u$, down $d$, and strange $s$, imply an approximate flavor $SU(3)$, which requires that baryons made of these quarks belong to the multiplets on the right side of the 'equation' $3\otimes 3\otimes 3 = 10_S \oplus 8_M \oplus 8_M \oplus 1_A$.Here the subscripts indicate symmetric, mixed symmetry, or anti-symmetric states under interchange of any two quarks [1].These were classified and arranged by Gell-Mann and Okubo in De-Couplets ($J$ = 3/2, $l$ = 0) and Octets ($J$ = 1/2, $l$ = 0) with +1 parities [2–4]. The Gell-Mann / Okubo mass formula which relates the masses of members of the baryon octet is given by; [2–4]

$$2(m_N + m_\Xi) = 3m_\Lambda + m_\Sigma \qquad (1)$$

While mass formula for De-couplets consists of equal spacing between the rows. The spaces are equal to an average value $-151\text{MeV/c}^2$.

$$M_\Delta - M_{\Sigma^*} = M_{\Sigma^*} - M_{\Xi^*} = M_{\Xi^*} - M_\Omega \qquad (2)$$

Gell-Mann used this formula and predicted the mass of the $\Omega^-$ baryon in 1962, equal to $M_\Omega = 1685 MeV/c^2$ [2]. Whereas the actual mass of the $\Omega^-$ is 1672 MeV/c$^2$ which was observed in 1964 [4]. Their mass difference is only 0.72 %, or in other words it was 99% true guess.

Now let's move towards baryon types made from the combination of four quarks, i.e. up *u*, down *d*, strange *s* and charm *c*. These belong to *SU*(4) multiplets. The *SU*(4) multiplets numerology is given by

$$4\otimes 4\otimes 4 = 20_S \oplus 20_M \oplus 20_M \oplus 4_A \ [5].$$

We are interested in twenty particles having spin 3/2 and even parity +1 forming one of the *SU*(4) multiplets. These particles are in their ground states, that is with $l = 0$. Their masses and other properties are listed in the Table 1.

Since the mass splitting due to strangeness is typically of order 151 $MeV/c^2$ in the de-couplet, which is an effect of about 13% of the masses it contains. The splitting due to charm in the *SU*(4) particles multiplets is expected to be larger [6]. The ratio $\frac{m_2}{m_1} = \frac{m_4}{m_3}$ indicates the direction of the increasing mass in the Strangeness, or Charm-ness planes in *SU*(3) and *SU*(4) multiplets respectively. It gives the mass symmetry breaking ratio for the multiplet [6]. In the *SU*(3) framework the Gell-Mann / Okubo relation for the $J^P = 1/2^+$ Octets, Eq. (1) and the equal spacing rule for the $J^P = 3/2^+$ de-couplets, Eq. (2) work so nicely that we cannot abandon linear mass formulae for baryons [6]. Using the above relation for baryons of $J^P = 3/2^+$ de-couplet of *SU*(3) gives a mass symmetry breaking ratio of order 1.0.

Table 1. List of the spin $J^P = 3/2^+$ particles formed from three quarks combinations (*qqq*) of four quarks *u*, *d*, *s* and *c* [5].

| **S.#** | **Particle Name** | **Symbol** | **Quark contents** | **Rest Mass ($MeV/c^2$)** |
|---|---|---|---|---|
| 1 | Delta ++ | $\Delta^{++}$ (1232) | *uuu* | 1232±2 |
| 2 | Delta Plus | $\Delta^{+}$ (1232) | *uud* | 1232±2 |
| 3 | Delta Zero | $\Delta^{0}$ (1232) | *udd* | 1232±2 |
| 4 | Delta Minus | $\Delta^{-}$ (1232) | *ddd* | 1232±2 |
| 5 | Sigma Plus | $\Sigma^{*+}$ (1385) | *uus* | 1382.8±0.4 |
| 6 | Sigma Zero | $\Sigma^{*0}$ (1385) | *uds* | 1383.7±1.0 |
| 7 | Sigma Minus | $\Sigma^{*-}$ (1385) | *dds* | 1387.2±0.5 |
| 8 | charmed Sigma ++ | $\Sigma_c^{*++}$ (2520) | *uuc* | 2517.9±0.6 |
| 9 | charmed Sigma + | $\Sigma_c^{*+}$ (2520) | *udc* | 2517.5±2.3 |
| 10 | charmed | $\Sigma_c^{*0}$ (2520) | *ddc* | 2518.8±0.6 |

| | Sigma Zero | | | |
|---|---|---|---|---|
| 11 | Xi Zero | $\Xi^{*0}$ (1530) | *uss* | 1531.80±0.32 |
| 12 | Xi Minus | $\Xi^{*-}$ (1530) | *dss* | 1535.0±0.6 |
| 13 | charmed Xi $^{+}$ | $\Xi_c^{*+}$ (2645) | *usc* | $2645.9^{+0.5}_{-0.6}$ |
| 14 | charmed Xi Zero | $\Xi_c^{*0}$ (2645) | *dsc* | 2645.9±0.5 |
| 15 | Omega Minus | $\Omega^{-}$ (1672) | *sss* | 1672.45±0.29 |
| 16 | charmed Omega Zero | $\Omega_c^{*0}$ (2770) | *ssc* | 2765.9±2.0 |
| 17 | double charmed Xi $^{++}$ | $\Xi_{cc}^{*++}$ | *ucc* | Unknown |
| 18 | double charmed Xi $^{+}$ | $\Xi_{cc}^{*+}$ | *dcc* | Unknown |
| 19 | double charmed Omega $^{+}$ | $\Omega_{cc}^{*+}$ | *scc* | Unknown |
| 20 | triple charmed Omega $^{++}$ | $\Omega_{ccc}^{*++}$ | *ccc* | Unknown |

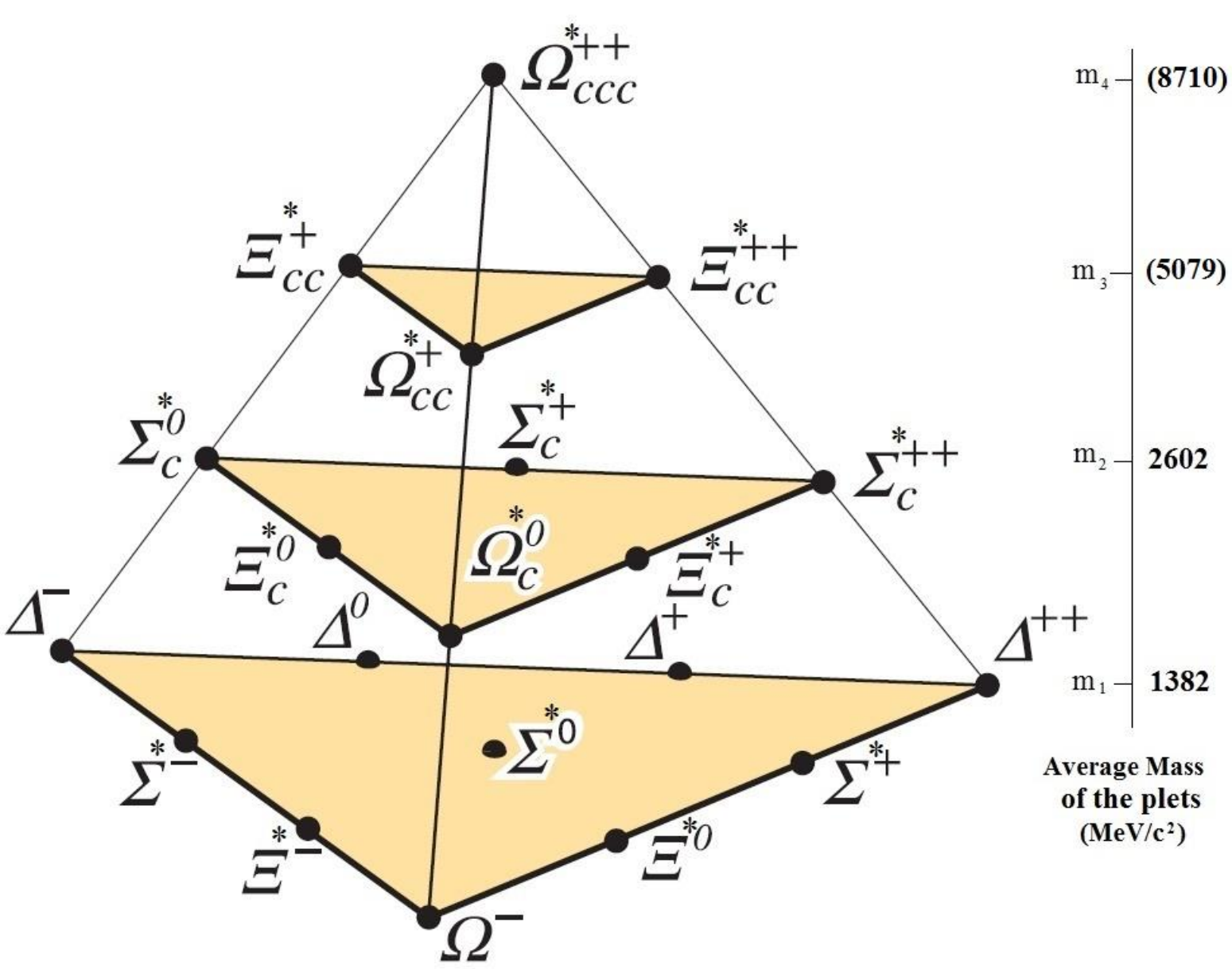

Fig. 1. *SU*(4) 20–plet of Baryons ($J^P$=3/2$^+$) made of *u*, *d*, *s* and *c* quarks, with an *SU*(3) de-couplet at the bottom [5].

## II. METHOD

There are four particles in the Table 1, whose masses are unknown. Let's try to make some predictions about the unknown masses of the $\Xi_{cc}^{*++}$ , $\Xi_{cc}^{*+}$ , $\Omega_{cc}^{*+}$ , and $\Omega_{ccc}^{*++}$ hyperons. Let the particles listed in Table 1 may be grouped in pyramid shape of increasing charm number from $C = 0$ to $C = 3$ as shown in Fig. 1 [5]. Masses of the particles are expressed in round numbers. The bottom of the 'Pyramid' is the $SU(3)$ de-couplet of $J^P = 3/2^+$ particles at $C = 0$. Average masses written on the right side of the Fig. 1 are discussed later in the text.

We can expand this 20–plet given in Fig. 1 in the shape as shown in Fig. 2, with increasing Charge Number from bottom to top. Also there becomes a beautiful pattern of Isospin numbers, $I$ of the particles. We can treat it as follows: Adding masses of the $\Delta^{++}$ and $\Delta^{+}$ in the first row of Charm Number $C = 0$, gives 2464 $MeV/c^2$, which is 54 less than the mass of the $\Sigma_c^{*+}$ (2518) particle in 2nd row of Charm Number $C = 1$. Hence approximately equal to its mass. Similarly Adding masses of $\Delta^{+}$ and $\Delta^{0}$ particles gives 2464 $MeV/c^2$, which is 55 less than the mass of the $\Sigma_c^{*0}$ (2519) particle in 2nd row of Charm Number $C = 1$. Same can be done for other particles. It gives approximate mass of the particle in the adjacent row with higher charm-ness.

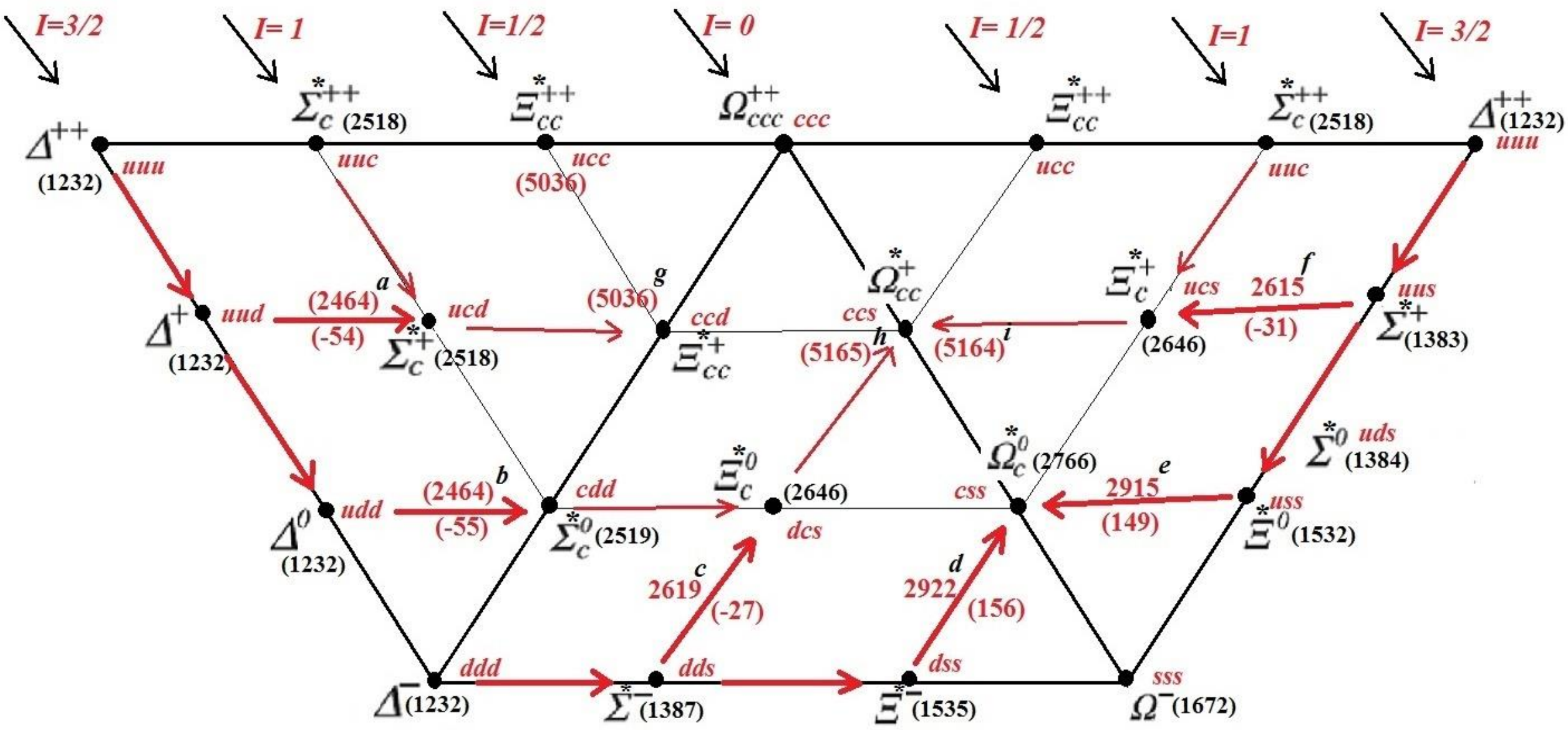

Fig. 2. $SU(4)$ 20–plet of Baryons ($J^P$=3/2$^+$) made of $u$, $d$, $s$ and $c$ quarks, arrows represent lines of the addition of particles masses. Indexed with $a$-$e$ (See Table 2), and numbers beneath the arrows are mass difference between theoretical and experimentally observed masses of the particles.

## II MASSES OF THE $\Xi_{cc}^{*+}$ AND $\Xi_{cc}^{*++}$

Assuming the same behavior of the masses, we can sum up masses of two particles in 2nd row of Charm Number $C = 1$, and get mass of the particle in 3rd row with charm Number $C = 2$.

Sum of the masses calculated in such a manner are indexed from (*a*) to (*i*) in Fig. 2; and are listed in Table 2. Masses of the Particles in 3rd row of charm Number $C = 2$ are unknown. Using this method the masses of these unknown particles $\Xi_{cc}^{+}$ and $\Omega_{cc}^{+}$ have been obtained and are equal to 5036 $MeV/c^2$ and 5164 $MeV/c^2$ respectively, as can be seen in Table 2. Since $\Xi_{cc}^{*+}$ and $\Xi_{cc}^{*++}$ are two different states of same particle, i.e. their Iso-spin is equal to ½, therefore both will approximately have equal masses, equal to 5036 $MeV/c^2$. But how could we guess that these two particles should have same iso-spin? Let us check.

### ISO-SPIN OF THE PARTICLES

If we see Fig. 2, values of Iso-spin numbers are decreasing from $I = +3/2$ to $I = 0$ by –½, from left to right, and then increasing by +½ till +3/2. Similarly, values of the third component of Iso-spin, $I_z$ or $I_3$, decreases for particles of same Iso-spin quantum number, at upper horizontal row to particles at lower horizontal row by 1. *SU*(4) 20–plet of Baryons ($J^P=3/2^+$) shown in figure 2 may be presented with details of $I$ and $I_3$, as shown in figure (3). Here iso-spin $I$ and its third component $I_3$ are presented in following figures as ($I$ , $I_3$).

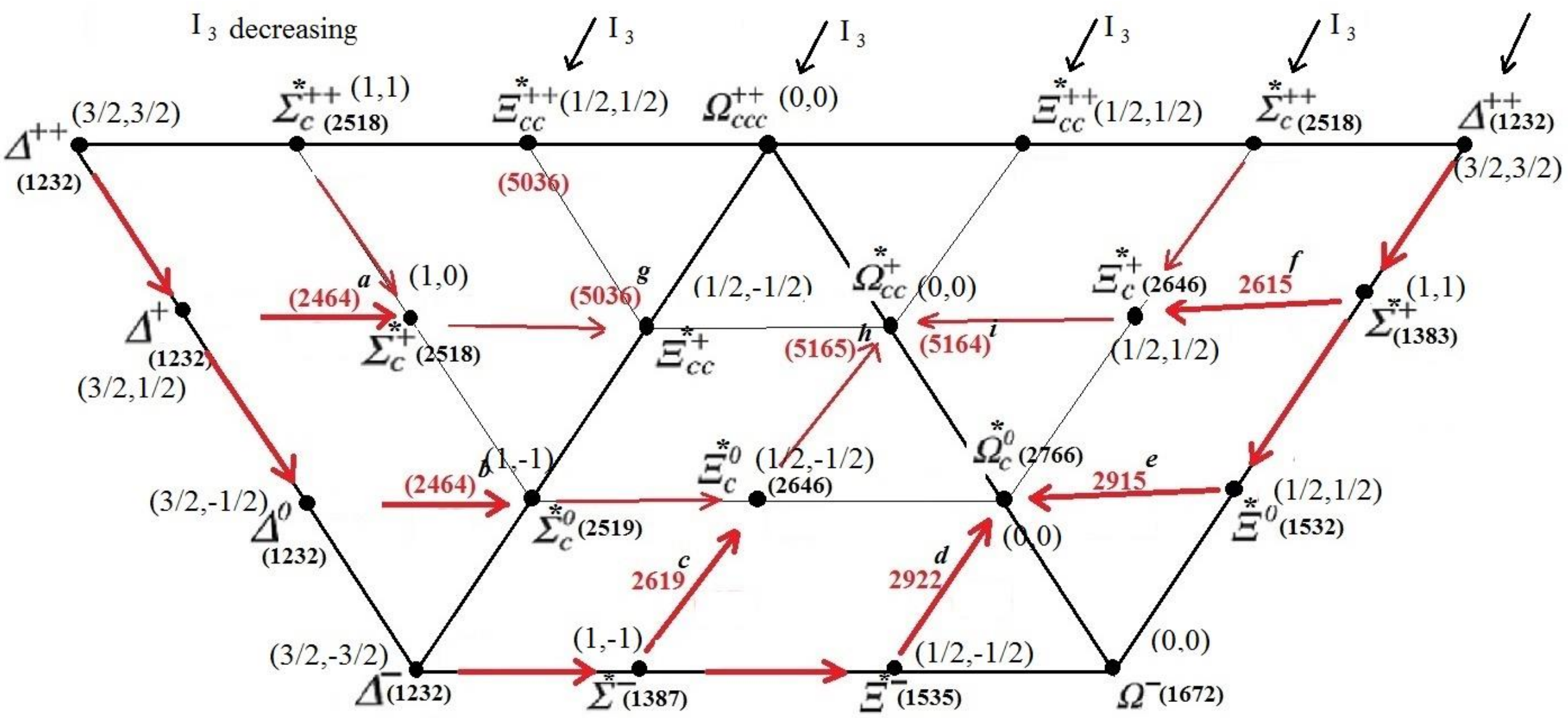

**Figure 3**. *SU*(4) 20–plet of Baryons ($J^P=3/2^+$) presented with details of $I$ and $I_3$. Here iso-spin $I$ and its third component $I_3$ are presented as ($I$ , $I_3$).

Here in this figure it may be observed that charge value in layers is increasing from bottom to top. Iso-spin has also a definite value in any diagonal layer (from top left side) from top to bottom. It is found from figure (2) and (3) that iso-spins of the undiscovered exotic hyperons $\Xi_{cc}^{*++}$, $\Xi_{cc}^{*+}$, $\Omega_{cc}^{*+}$ and $\Omega_{ccc}^{++}$ should be 1/2 , 1/2, 0, and 0 respectively. Similarly in order to find third component of iso-spin ($I_3$) of these particles, in figure (3) it may be observed that $I_3$ of particles in diagonal layers (from top right side) is decreasing by 1/2 for each layer from top to bottom. By setting the values of $I$ and $I_3$ for undiscovered hyperons according to the above procedure, it is found that $\Xi_{cc}^{*++}$, $\Xi_{cc}^{*+}$, $\Omega_{cc}^{*+}$ and $\Omega_{ccc}^{++}$ should have values of $I_3$ equal to 1/2, -1/2, 0 and 0 respectively.

Table 2. Calculations of masses of the particles from sum of other particles, indexed in Fig. 2.

| S# | Index | Particle 1 ($m_1$) | Particle 2 ($m_2$) | $m_1 + m_2$ (MeV/$c^2$) | Resultant Particle (m) | $(m_1+m_2)-m$ |
|---|---|---|---|---|---|---|
| 1 | *a* | $\Delta^{++}$ (1232) | $\Delta^{+}$ (1232) | 2464 | $\Sigma_c^{*+}$ (2518) | -54 |
| 2 | *b* | $\Delta^{+}$ (1232) | $\Delta^{0}$ (1232) | 2464 | $\Sigma_c^{*0}$ (2519) | -55 |
| 3 | *c* | $\Delta^{-}$ (1232) | $\Sigma^{*-}$ (1387) | 2619 | $\Xi_c^{*0}$ (2646) | -27 |
| 4 | *d* | $\Sigma^{*-}$ (1387) | $\Xi^{*-}$ (1535) | 2922 | $\Omega_c^{*0}$ (2766) | 156 |
| 5 | *e* | $\Sigma^{*+}$ (1383) | $\Xi^{*0}$ (1532) | 2915 | $\Omega_c^{*0}$ (2766) | 149 |
| 6 | *f* | $\Delta^{++}$ (1232) | $\Sigma^{*+}$ (1383) | 2615 | $\Xi_c^{*+}$ (2646) | -31 |
| 7 | *g* | $\Sigma_c^{*++}$ (2518) | $\Sigma_c^{*+}$ (2518) | 5036 | $\Xi_{cc}^{*+}$ (??) | ?? |
| 8 | *h* | $\Sigma_c^{*0}$ (2519) | $\Xi_c^{*0}$ (2646) | 5165 | $\Omega_{cc}^{*+}$ (??) | ?? |
| 9 | *i* | $\Sigma_c^{*++}$ (2518) | $\Xi_c^{*+}$ (2646) | 5164 | $\Omega_{cc}^{*+}$ (??) | ?? |

Mass of the double charmed *Xi* ++, $\Xi_{cc}^{++}$ (probably with $J^P = ½^+$) is approximately equal to 3621 *MeV/c*$^2$ [6]. We can compare this value with the theoretical mass of the $\Xi_{cc}^{*++}$ particle with $J^P = 3/2^+$, we extracted i.e. 5036 *MeV/c*$^2$. Percentage difference between these two masses is $\frac{5036-3621}{3621}$, which is equal to 39%. It is in agreement with the percentage mass difference between masses of proton/ neutron and delta resonances (having same quark contents, but different spins) $\frac{1232-938}{938}$ equal to 31%. It is acceptable, and hence we can take 5036 MeV/$c^2$ as approximate mass of the $\Xi_{cc}^{*+}$ and $\Xi_{cc}^{*++}$ particles with some uncertainty equal to its square root, given by ± 71 MeV/$c^2$.

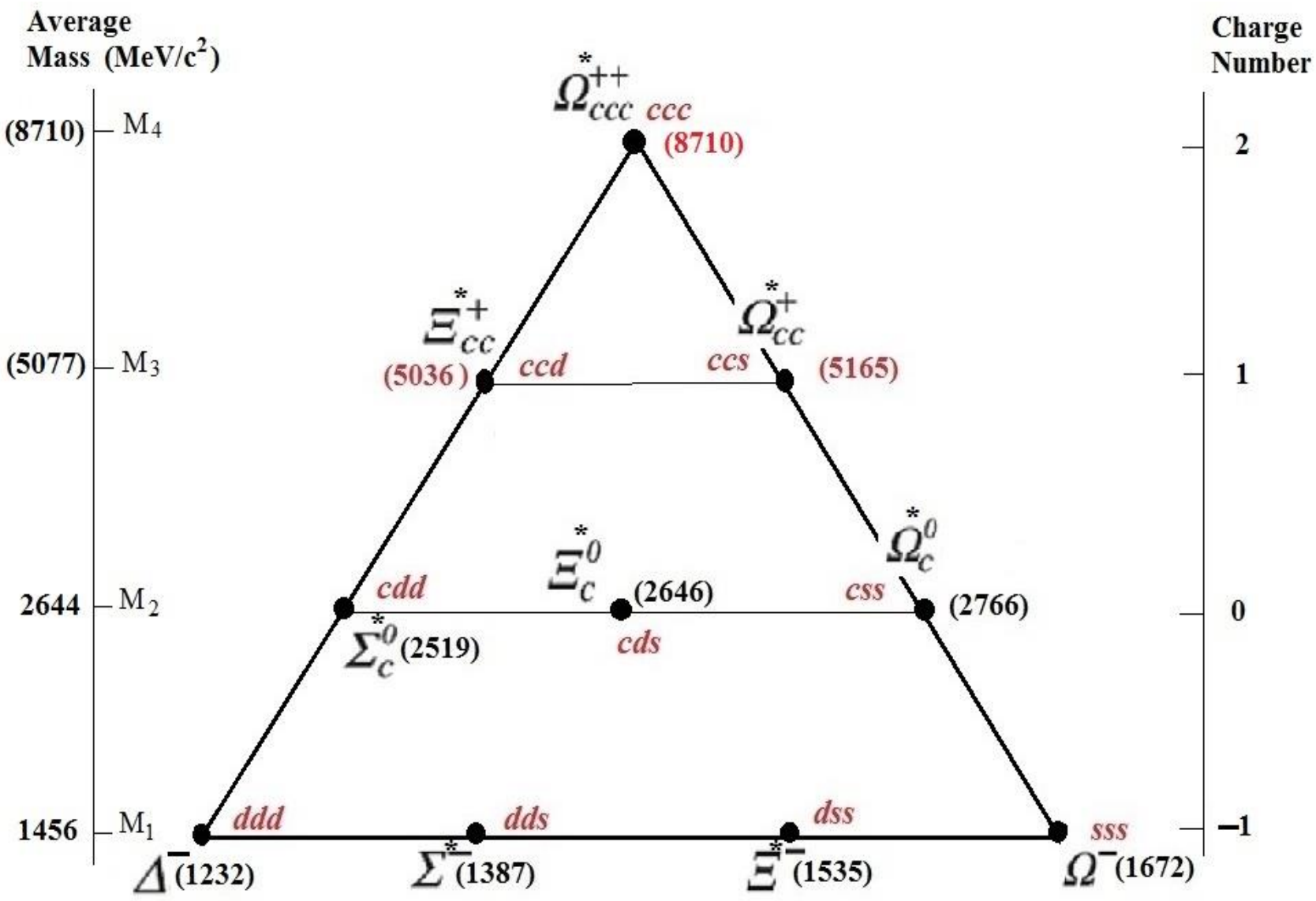

**Figure 4.** A portion of the SU(4) Multiplets, with increasing Mass, Charge number and Charmness from bottom to top.

## ANOTHER TEST

Mass of the $\Omega^{+}_{cc}$ hyperon having $(J)^{P}=(1/2)^{+}$ has been predicted [7] to be 3916±62 $MeV/c^2$. Percentage difference between particles with same quark contents but different total spin in average, calculated above (0.39 and 0.31, is equal to 0.35. The mass of the $\Omega^{+*}_{cc}$ having $(J)^{P}=(3/2)^{+}$ should be:

$$3916\times 0.35(3916)=5286\,MeV/c^{2}\approx 5165\pm 72 MeV/c^{2}$$

Or if we consider it 0.33, it gives exact mass of the $\Omega^{+*}_{cc}$ particle:

$$3916\times 0.33(3916)=5208\,MeV/c^{2}\approx 5165\pm 72 MeV/c^{2}$$

## III MASS OF THE $\Omega^{++}_{ccc}$ PARTICLE

Mass of the $\Omega^{++}_{ccc}$ particle can be estimated using Fig. 4. Taking average mass value of first three rows i.e. rows with $C$ = 0, 1, 2 and finding their difference may lead us to the mass of the particle at the top, with $C$ = 3.

$M_1$ (Average mass of row with $C$ = 0) = 1456 $MeV/c^2$

$M_2$ (Average mass of row with $C$ = 1) = 2644 $MeV/c^2$

$M_3$ (Average mass of row with $C$ = 2) = 5100 $MeV/c^2$

$M_4$ = mass of $\Omega^{++}_{ccc}$ particle.

Table 3. Approximated masses and other properties of the unknown particles in Standard Model of $SU(4)$.

| S.# | Particle | Mass ($MeV/c^2$) | Iso spin | | Parity ($P$) | $I(J)^P$ |
|---|---|---|---|---|---|---|
| | | | $I$ | $I_z$ | | |
| 1 | $\Xi_{cc}^{*++}$ | $5036 \pm 71$ | $\frac{1}{2}$ | $+\frac{1}{2}$ | +1 | $\frac{1}{2}(\frac{3}{2})^+$ |
| 2 | $\Xi_{cc}^{*+}$ | $5036 \pm 71$ | $\frac{1}{2}$ | $-\frac{1}{2}$ | +1 | $\frac{1}{2}(\frac{3}{2})^+$ |
| 3 | $\Omega_{cc}^{*+}$ | $5165 \pm 72$ | 0 | 0 | +1 | $0(\frac{3}{2})^+$ |
| 4 | $\Omega_{ccc}^{++}$ | $8710 \pm 93$ | 0 | 0 | +1 | $0(\frac{3}{2})^+$ |

Difference between $M_2$ and $M_1$:

$$M_2 - M_1 = 2644 - 1456 = 1188\, MeV/c^2$$

Similarly;

$$M_3 - M_2 = 5100 - 2644 = 2456\, MeV/c^2$$

$$\text{Also} \quad 1188 \times 2 = 2376\, MeV/c^2,$$

which is very near to 2456 $MeV/c^2$ (with only difference of 80 $MeV/c^2$). It means;

$$M_2 - M_1 = 1188\, MeV/c^2$$

$$M_3 - M_2 = 2(M_2 - M_1) + 80 = (2 \times 1188) + 80\ MeV/c^2$$

and

$$M_4 - M_3 = 3(M_2 - M_1) + 46 = (3 \times 1188) + 46\, MeV/c^2$$

Now we can generalize these in one equation:

$$\boxed{M_n = M_{n-1} + Charmness \times (1188) + Charge \times (\sqrt{1188})} \qquad (3)$$

Where n=2, 3, 4

That is;

$$M_2 = M_1 + 1 \times (1188) + 0 \times (\sqrt{1188})$$
$$= 1456 + 1188 = 2644\, MeV/c^2$$

$$M_3 = M_2 + 2 \times (1188) + 1 \times (\sqrt{1188})$$
$$= 2644 + 2376 + 34 = 5054\, MeV/c^2$$

Here value of $M_3$ will be average of both the theoretical values, as there are no experimentally observed values for particles in this row. i.e.

$$M_3 = \frac{5100 + 5054}{2} = 5077\, MeV/c^2$$

hence

$$M_4 = M_3 + 3 \times (1188) + 2 \times (\sqrt{1188})$$
$$= 5077 + 3564 + 69 = 8710\, MeV/c^2$$

Thus the mass of $\Omega_{ccc}^{++}$ particle, estimated using the above method is equal to 8710 ± 93 $MeV/c^2$. Statistical Errors are the square root values of the particle masses. Experimentally observed Particle masses may be slightly differ from the above estimated values of masses of all the four particles, due to rough estimates. However in that case the values obtained experimentally should agree with theoretical mass values with in two standard errors. All the predicted masses of the four unknown particles and their other properties are listed in Table 3.

Now let us check what is the mass difference, $M_2 - M_1 = 1188\, MeV/c^2$. It is actually mass of the Charm quark, which is the constituent of the particles. Charm number increases by +1 as we go up in the particles pyramid of Fig. 1, and Fig. 4. So this increase in the masses of the particles is due to the mass of the charm quark. Comparing this with the evaluated mass of the charm quark $M_c = 1270 \pm 30$ and $M_c = 1196 \pm 59 \pm 50$ $MeV/c^2$ listed in Ref. [5] gives satisfactory result. Hence it is also proved that splitting due to charm quarks in the $SU(4)$ particles multiplets is about an order of magnitude, equal to an average value of 80%, as was expected [8].

The calculations may also be expanded to the whole 20- multiplet, as shown in Fig. 1, so that a combined formula for all the baryons masses of this multiplet may be extracted.

In the Fig. 1, $m_1$ shows average mass of the particles forming de-couplet, with $C = 0$, and is equal to 1382 $MeV/c^2$, $m_2$ = 2602 $MeV/c^2$ is average mass of the particles with $C = 1$, $m_3$ = 5079 $MeV/c^2$ is the average mass of particles with $C = 2$ and $m_4$ = 8710 $MeV/c^2$ is the average mass of particle having $C = 3$.

General expression for these masses may be derived as under;

$$m'_n = m_{n-1} + Charmness \times (1188) + Charmness \times (\sqrt{1188})$$

Where n=2, 3, 4

Or more generally,

$$\boxed{m'_n = m_{n-1} + Charmness \times (1218)} \qquad (4)$$

with n=2, 3, 4

Let's try to find average mass of the particles multiplets using this relation;

$$m'_2 = m_1 + C \times (1218) = 1382 + 1 \times (1218)$$
$$= 2600\, MeV/c^2 \approx m_2$$

$$m_3' = m_2 + C \times (1218) = 2602 + 2(1218)$$
$$= 5038\, MeV/c^2 \approx m_3$$

and

$$m_4' = m_3 + C \times (1218) = 5079 + 3(1218)$$
$$= 8733\, MeV/c^2 \approx m_4$$

Or equation (4) can also be written as under:

$$\boxed{m_n' - m_{n-1} = 1218\,(n-1)\, MeV/c^2} \qquad (5)$$

with n = 2, 3, 4

Mass difference term 1218 $MeV/c^2$ is also equal to the mass of the charm quark [5]. The mass symmetry braking ratio $\frac{m_2}{m_1} = \frac{m_4}{m_3}$ in the direction of the increasing mass in the Charm plane for $SU(4)$ multiplets, is of the order of an average value of 1.8 for baryons of $J^P=3/2^+$.

## III. Conclusions

$SU(4)$ multiplets of Standard Model have been analyzed in this article and some predictions have been made about the unknown masses and other properties of the $\Xi_{cc}^{*++}$ , $\Xi_{cc}^{*+}$ , $\Omega_{cc}^{*+}$ , and $\Omega_{ccc}^{++}$ hyperons. Their masses are estimated to be $\Xi_{cc}^{*++}$ (5036 ± 71), $\Xi_{cc}^{*+}$ (5036 ± 71), $\Omega_{cc}^{*+}$ (5165 ± 72), and $\Omega_{ccc}^{++}$ (8710 ± 93) $MeV/c^2$. Statistical errors are the square root values of the estimated masses of particles. Experimentally observed Particles' masses may slightly differ from the estimated values of masses of all the four particles, due to rough estimates. However in that case the values obtained experimentally should agree with theoretical mass values with in two standard errors. The mass symmetry braking ratio for $SU(4)$ multiplets, is estimated to be of the order of an average value of 1.8 for baryons of $J^P=3/2^+$. It is found that $\Xi_{cc}^{*++}$ and $\Xi_{cc}^{*+}$ have same isospin, equal to 1/2, so both would have approximately equal masses. $\Xi_{cc}^{++}$ has third component of isospin $I_z$ equal to +1/2, while that of $\Xi_{cc}^{*+}$ is equal to –1/2. While other two particles have $I$ and $I_3$ equal to zero. A general formula for masses has been derived for the $SU(4)$ multiplets, which gives satisfactory results for observed and estimated masses of the particles of $SU(4)$. The mass difference in the multiplets of the $SU(4)$ is due to the mass of Charm quark, and its value is also equal to mass of the charm quark. The results in this research, if proved correct, will be helpful in understanding the particle physics, and will strengthen the Standard Model of particles.

**ACKNOWLEDGMENTS**

Thanks are due to Dr. Khusniddin Olimov who is my Ph.D Supervisor and to Dr. Farida Tahir, who introduced me with Particle Physics and inspired me during a lecture to the need for some mass formula and predictions about masses of missing hyperons of the $SU(4)$ multiplets.